%% file: main.tex
\documentclass{article}
\usepackage{spconf,amsmath,graphicx,tikz,calc,enumitem,subcaption,algorithm,algpseudocode,numprint,comment}
\usepackage[keeplastbox]{flushend}
\usetikzlibrary{calc}

\usepackage[hyphens,spaces,obeyspaces]{url}
\usepackage[breaklinks]{hyperref}
\hypersetup{breaklinks=true}
\usepackage{cite}
\usepackage{url}
\usepackage{ragged2e}
\usepackage{amssymb}
\usepackage{siunitx}

\usepackage{flushend} 
\usepackage{booktabs}
\usepackage{cleveref}
\usepackage{bm}
\usepackage{multirow} 
\usepackage{array}
\usepackage{booktabs}
\usepackage{mathrsfs}
\newcolumntype{N}{>{\centering\arraybackslash}m{0.8cm}}
\crefname{figure}{Fig.}{}
\crefname{table}{Table}{}
\crefname{section}{Section}{}
\crefname{equation}{Eq.}{}

\newcommand\rightlast{\leftskip0ptplus1fil\rightskip0ptplus-1fil\parfillskip0ptplus1fil}\DeclareCaptionJustification{rightlast}{\rightlast}

\ninept

\npfourdigitnosep 

\algrenewcommand\textproc{}

\title{prediction of object geometry from acoustic scattering using convolutional neural networks}

\name{Ziqi Fan$^1$, Vibhav Vineet$^2$, Chenshen Lu$^1$, T.W. Wu$^{3}$, Kyla McMullen$^1$}
\address{$^1$University of Florida, Gainesville, USA\\
$^2$Microsoft Research, Redmond, USA\\
$^3$University of Kentucky, Lexington, USA}
  
\begin{document}
\maketitle

\begin{abstract}
Acoustic scattering is strongly influenced by boundary geometry of objects over which sound scatters. The present work proposes a method to infer object geometry from scattering features by training convolutional neural networks. The training data is generated from a fast numerical solver developed on CUDA. The complete set of simulations is sampled to generate multiple datasets containing different amounts of channels and diverse image resolutions. The robustness of our approach in response to data degradation is evaluated by comparing the performance of networks trained using the datasets with varying levels of data degradation. The present work has found that the predictions made from our models match ground truth with high accuracy. In addition, accuracy does not degrade when fewer data channels or lower resolutions are used.

\end{abstract}


\input{text/Introduction.tex}

\input{text/Formulation.tex}
\input{text/Experiments.tex}
\input{text/Conclusion.tex}




\bibliographystyle{IEEEtran}
\bibliography{Bib/DeepLearning,Bib/RoomGeoEst,Bib/Sonar,Bib/books}

\end{document}

%% file: text/Introduction.tex
\section{Introduction}
Acoustic reflection models have been extensively researched to predict room geometry\cite{Antonacci2010Reconstruction,Ba2010L1,Baba2016Multiple,Canclini2011Doa2Loc,Canclini2011Exact,Dokmanic2013NAS,Dokmanic2011RoomShape2D,Filos2012Linear,Filos2010TwoStep,Michael2019Spherical,Moore2013SingleChannel,Naseri2017Cooperative,Nastasia2011Localization,Remaggi2014Room,Remaggi2017Novel,Remaggi20153D,Remaggi2015Source,Tervo20123D,Antonacci2012InferenceRoom,Dokmanic2016RoomSlam}. A widely-used approach is based on the assumption that walls can be described by parametric lines in a 2-D plane \cite{Antonacci2010Reconstruction,Baba2016Multiple,Canclini2011Doa2Loc,Canclini2011Exact,Filos2010TwoStep,Remaggi2014Room,Remaggi20153D,Antonacci2012InferenceRoom}. In this approach, ellipses with source-receiver pairs as their focal points are constructed. After a sonic event, acoustic measurements such as time of arrival (TOA) \cite{Antonacci2010Reconstruction,Antonacci2012InferenceRoom,Baba2016Multiple,Canclini2011Exact,Filos2010TwoStep,Remaggi2014Room,Remaggi20153D} and direction of arrival (DOA) \cite{Canclini2011Doa2Loc,Remaggi2014Room} are extracted from the room reflections and used to determine shapes of the ellipses. The parametric lines forming the walls are determined from common tangential lines of the ellipses. In another popular approach, the virtual image sources mirrored from real sources are used to determine the orientations of the walls \cite{Dokmanic2016RoomSlam,Dokmanic2011RoomShape2D,Dokmanic2013NAS,Moore2013SingleChannel,Michael2019Spherical,Naseri2017Cooperative,Nastasia2011Localization,Remaggi2017Novel,Remaggi2015Source}. In this approach, to determine the wall orientation, a series of circles centered at different receiver locations are constructed. The radii of these circles are determined by the TOAs of echos from the wall. The image source is at the intersection of the circles and the wall is perpendicular to the line segment connecting the image source and the real source.

Recently, Lindell et al. \cite{Lindell2019Nlos} proposed predicting the shape of an object hidden behind a corner using measurements of reflections. In their work, acoustic waves experience \numprint{3} reflections before finally arriving at a microphone array. A computational model is derived for the mapping from received acoustic waves to surface geometry. The object surface is reconstructed from the microphone-recorded signals. Their work shows that reflection models are able to predict object geometry in more complicated environments. 

The previously mentioned studies only use reflection, despite the many other main effects of acoustic scattering. Critical features such as diffraction and occlusion are neglected. Recently, Fan et al. \cite{Fan2019Fast} proposed to predict acoustic scattering from objects using a convolutional neural network (CNN). The CNN was trained using images representing object geometry and acoustic fields. Reflections, diffractions, and occlusions were accurately constructed from the CNN, according to their test cases. Inspired by \cite{Fan2019Fast}, it is believed that the inverse problem, prediction of object geometry from acoustic scattering using a CNN, is also feasible.

\begin{figure*}[t!]
\centering
\includegraphics{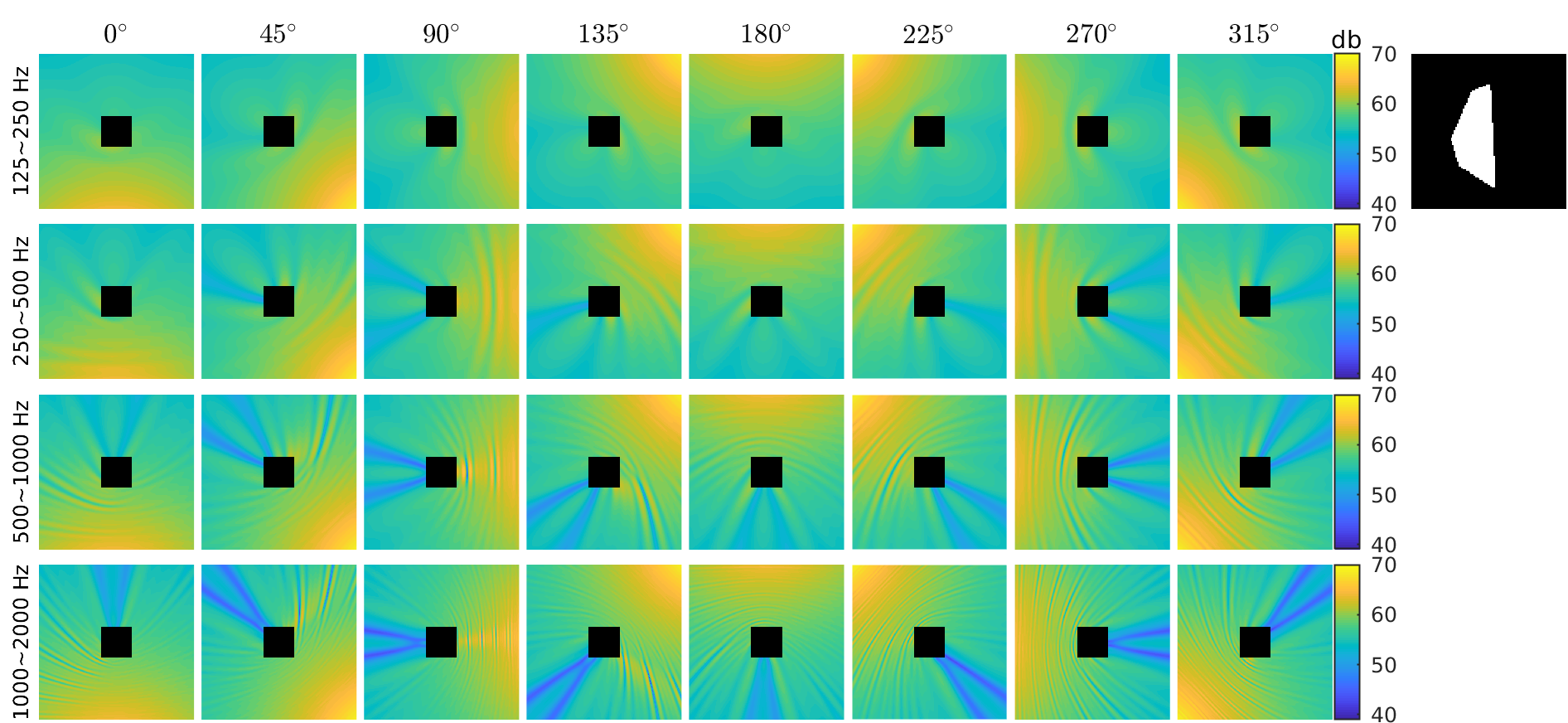}
\caption{An input-output pair to the neural network of the largest possible number of channels and the highest resolution. The input is a \numprint{32}-channel image, \numprint{4} channels for octave bands and \numprint{8} channels for source locations. Bright yellow pixels far from the dark region denote locations near sound sources. Bright yellow curves radiating from the dark region denote reflection from the object. Dark blue streaks denote occlusion of sound propagation. Green lobes between streaks of occlusion denote diffraction around the object. The output is a binary image indicating the geometry of an object in the inaccessible region. White and black pixels denote object occupancy and air respectively.}
\label{fig: input-output pair}
\includegraphics{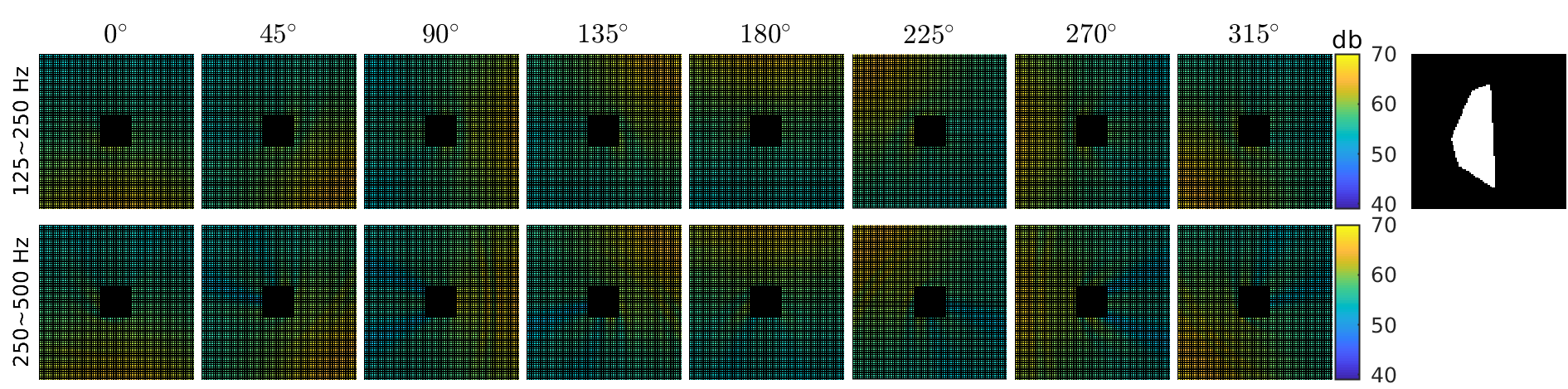}
\caption{An input-output pair using low-frequency octave bands only and a spatial sampling factor of \numprint{2}. Black pixels denote unknown loudness. A much smaller number of measurements is needed for this input, as compared to that in \cref{fig: input-output pair}. The number of sources can be reduced to generate images of fewer source channels. Density of virtual probes can be further decreased to lower image resolution.}
\label{fig: decimated input-output pair}
\vspace{-12pt}
\end{figure*}

The present work presents the first study that predicts object geometry from acoustic scattering using CNNs. To achieve this goal, we developed a fast acoustic solver on an Nvidia GPU to generate a large amount of training pairs. An example of these pairs is shown in \cref{fig: input-output pair}. In reality, collecting data shown in \cref{fig: input-output pair} is expensive, so we aim to reduce the number of channels and image resolution for a more practical solution. Through selecting multiple source locations, differentiating low and high frequency ranges, and varying density of virtual sensors, we created \numprint{24} datasets from the original data and a separate CNN was trained for each of them. The \numprint{24} CNN models were carefully studied to evaluate the influence of data degradation on prediction accuracy.

Our work shows that synthetic data can be used with CNN models to accurately predict object geometry. Further, using degraded acoustic images does not significantly decrease accuracy. The results indicate that predicting object geometry from acoustic scattering using CNNs is a robust and promising approach. Our training and test data and the numerical solver are shared at the following address:  \url{https://faculty.eng.ufl.edu/soundpad-lab/research/aml/}.

%% file: text/Formulation.tex
\section{Methodology}
\subsection{Problem formulation}
Let $\Omega$ denote a space of interest and $\mathcal{R}$ be a centered subspace of $\Omega$. There exist objects in $\mathcal{R}$. In this initial study, we restrict the number of objects to \numprint{1}. Let $\mathcal{V}$ and $\partial\mathcal{V}$ denote the volume and boundary of this object. Region $\mathcal{R}$ is inaccessible and information of $\partial\mathcal{V}$ is unknown. Let $\mathcal{S} = \Omega\setminus\mathcal{R}$ be an open region free of objects. A sound source can be placed at any location $\bm{x}_s\in\mathcal{S}$. Our task is to predict $\partial\mathcal{V}\subseteq\mathcal{R}$ using acoustic information in $\mathcal{S}$ induced by sources and $\partial\mathcal{V}$.

Pressure field $p$ induced by a point source at $\bm{x}_s$ at an angular frequency $\omega$ is described by the Helmholtz equation:
\begin{equation}
\nabla^2p\left(\omega,\boldsymbol{x};\bm{x}_{s}\right)+k^2p\left(\omega,\boldsymbol{x};\bm{x}_{s}\right) = -\delta\left(\boldsymbol{x}-\boldsymbol{x}_{s}\right),\ \bm{x}\in\Omega-\mathcal{V},
\label{eq: helmholtz equation}
\end{equation}
where $k = \omega/c$ denotes the wave number and $\delta$ denotes the Dirac delta. A unique solution of \cref{eq: helmholtz equation} is ensured by the Sommerfeld radiation condition at infinity \cite{williams1999fourier} and a frequency-dependent boundary condition defined on $\partial\mathcal{V}$:
\begin{equation}
\alpha\left(\omega,\bm{x}\right)p\left(\omega,\bm{x}\right) + \beta\left(\omega,\bm{x}\right)q\left(\omega,\bm{x}\right) = \gamma\left(\omega,\bm{x}\right),\ \bm{x}\in\partial\mathcal{V},
\label{eq: boundary condition}
\end{equation}
where $q = \partial p/\partial n\left(\bm{x}\right)$ denotes the normal velocity on $\partial\mathcal{V}$. Finally, we define loudness at a point $\bm{x}$ as spectral energy concentration:
\begin{equation}
L_{i,j}\left(\boldsymbol{x}\right) = 10\log_{10} \frac{1}{\omega_{i+1}-\omega_i}\int_{\omega_i}^{\omega_{i+1}}\lvert p\left(\omega,\boldsymbol{x};{\bm{x}_{s,j}}\right)\rvert^2\mathrm{d}\omega,\ \boldsymbol{x}\in\mathcal{S},
\label{eq: loudness}
\end{equation}
where $\bm{\omega}_{i} = \left[\omega_{i},\omega_{i+1}\right)$ denotes the $i$-th frequency range and $\bm{x}_{s,j}$ denotes the $j$-th source location.

Loudness fields $L$ and boundary $\partial\mathcal{V}$ are related by \cref{eq: boundary condition}. Two boundary properties, geometry $\mathcal{G}$ and acoustic impedance $\mathcal{Z}$ are associated with $\partial\mathcal{V}$. In this study, we assume $\partial\mathcal{V}$ to be acoustically rigid, corresponding to the case $\alpha = \gamma = 0$, $\beta \neq 0$ in \cref{eq: boundary condition}. Thus, the only variable related to $\partial\mathcal{V}$ is its geometry $\mathcal{G}$. A total number of $N_s$ source locations $\bm{x}_{s,j}$ uniformly distributed around an object and $N_f$ frequency ranges $\bm{\omega}_{i}$ in the form of octave bands are used. For each $\mathcal{G}$, $N_s\cdot N_f$ loudness fields are defined on $\mathcal{S}$, each loudness field corresponding to a unique combination of $\bm{\omega}_{i}$ and $\bm{x}_{s,j}$. The mapping $\mathcal{M}: \left\{L\right\} \to \left\{\mathcal{G}\right\}$ can be constructed with a large number of data pairs $\left(L,\mathcal{G}\right)$. Once $\mathcal{M}$ is constructed, prediction of an unknown $\mathcal{G}$ can be conveniently made if $L$ measurements are provided.

\subsection{Data representation}
In this study, data pairs $\left(L,\mathcal{G}\right)$ are provided by simulations. Fan et al. proposed a parallel acoustic solver developed in CUDA \cite{fan2019computation}. This solver is based on GPU implementation of the boundary element method \cite{wu2002boundary} and provides a fast estimation of pressure distributions on $\partial\mathcal{V}$ through solving \cref{eq: helmholtz equation}. The solver was further developed to extrapolate $p$ on $\partial\mathcal{V}$ into $\Omega \setminus \mathcal{V}$ and to calculate $L$ of \cref{eq: loudness} in $\mathcal{S}$.

Both $L$ and $\mathcal{G}$ are represented using voxel grids. Similar to \cite{raghuvanshi2018triton}, space $\Omega$ is divided into many small cubes and these cubes form a voxel grid $\mathbf{G}\left(\Omega\right) = \mathbf{G}\left(\mathcal{R}\right)\cup\mathbf{G}\left(\mathcal{S}\right)$. Loudness samples $L_{i,j}\left(\bm{x}\right)$ are calculated at every center point of a voxel in $\mathbf{G}\left(\mathcal{S}\right)$. Consequently, each voxel in $\mathbf{G}\left(\mathcal{S}\right)$ is associated with a numerical value in decibels. Voxels in $\mathbf{G}\left(\mathcal{R}\right)$ are associated with a special value denoting unknown, as $\mathcal{R}$ is inaccessible. All the loudness grids form an image $\mathcal{I}\left(L\right)$ of $N_f\cdot N_s$ channels, each channel denoting one combination of a source location and a frequency range. The geometric distribution of the object in $\Omega$ is also defined on $\mathbf{G}\left(\Omega\right)$. Each voxel of $\mathbf{G}\left(\Omega\right)$ is associated with a binary value, \numprint{1} denoting object occupancy and \numprint{0} denoting air. This association leads to a binary image $\mathcal{I}\left(\mathcal{G}\right)$ representing the geometry of the object. As objects are originally represented using mesh models composed of triangular elements, object occupancy is determined from triangle-cube intersection tests.

As a result, both $L$ and $\mathcal{G}$ of an object are represented using 3-D images $\mathcal{I}\left(L\right)$ and $\mathcal{I}\left(\mathcal{G}\right)$. The mapping $\mathcal{M}: \left\{L\right\} \to \left\{\mathcal{G}\right\}$ is then turned into a new mapping $\mathcal{M}^\prime: \left\{\mathcal{I}\left(L\right)\right\} \to \left\{\mathcal{I}\left(\mathcal{G}\right)\right\}$. CNNs are a good candidate for constructing $\mathcal{M}^{\prime}$ in the form of images. Our formulation is image segmentation in nature, where each pixel in an output image is of two categories: occupied by an object or not.

%% file: text/Experiments.tex
\section{experimental evaluation}
\subsection{Simulation setup}
\begin{figure}[t!]
\centering
\includegraphics{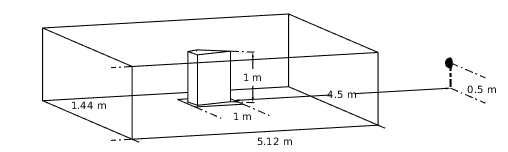}
\caption{Illustration of our simulation region. Only \numprint{1} out of \numprint{8} potential source locations around the region is shown.}
\label{fig: simulation region}
\end{figure}

The simulation region shown in \cref{fig: simulation region} is \SI{5.12}{\m}$\times$\SI{5.12}{\m}$\times$\SI{1.44}{\m}. This region is turned into a voxel grid using unit cubes of volume \SI{1}{\cm^3}. There is a \SI{1}{\m}$\times$\SI{1}{\m} square region at the center of the simulation region. An object whose geometry is to be determined resides within the square region. The distance from a sound source to the center of the region is fixed at $r_s = \SI{5}{\m}$. The sound source can be positioned at any of the following $N_s = 8$ directions: 
\begin{equation}
\theta_j = j\cdot\pi/4,\ j = 0,1,2,\dots,7.
\end{equation}
We also use $N_f = 4$ frequency ranges in the form of octave bands:
\begin{equation}
\bm{\omega}_i = \left[\omega_{i},\omega_{i+1}\right),\ \omega_i = 2\pi\cdot 125\cdot 2^{i+1},\ i=0,1,2,3.
\label{eq: octave bands}
\end{equation}

\subsection{Generation of objects and input-output pairs}

\begin{figure}[t!]
\centering
\includegraphics{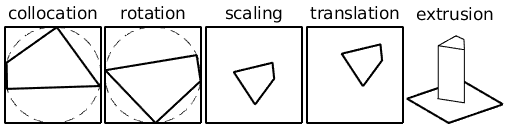}
\caption{Our pipeline to generate convex objects of diverse geometry, adapted from \cite{Fan2019Fast}. A translation step is added to ensure diverse locations. Objects from this pipeline only vary horizontally.}
\label{fig: pipeline}
\vspace*{\floatsep}
\includegraphics[width=0.45\textwidth]{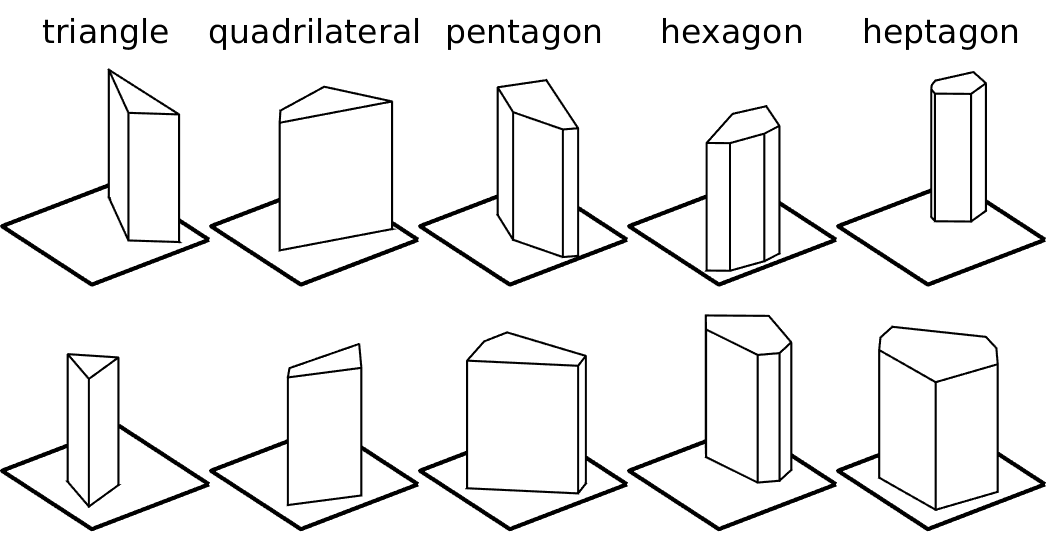}
\caption{Examples of objects for our training dataset. The objects cover \numprint{5} categories, each one extruded from a type of polygons.}
\label{fig: generated}
\end{figure}

We make the assumption that objects to be predicted are horizontally convex. We designed the pipeline shown in \cref{fig: pipeline} for object generation. The pipeline is adapted from \cite{Fan2019Fast}. Different from \cite{Fan2019Fast}, a step of translation is inserted before extrusion to ensure the diversity of object positions within the square region. We generated five categories of prisms, namely, triangular, quadrilateral, pentagonal, hexagonal and heptagonal to simulate training data. There are $N_{tc} = 3150$ instances in each category, summing up to a total number of $N_{tr} = 15750$. Examples of these objects are shown in \cref{fig: generated}. We also generated a smaller dataset for generalization test. Instead of collocating vertices on the circle inscribed to the square, we collocated vertices on circles of \numprint{4} different radii: \SI{0.3}{\m}, \SI{0.35}{\m}, \SI{0.4}{\m} and \SI{0.45}{\m}. The step scaling was not used in generating objects of the test dataset. For each radius, we generated the same \numprint{5} categories of prisms, each of \numprint{35} instances, summing up to $N_{tst} = 700$. We compared our test objects to training objects to ensure no overlap.

We simulated loudness fields $L$ for all objects in the training set and the test set. For each object, combinations from $N_s = 8$ source locations and $N_f = 4$ octave bands lead to \numprint{32} independent simulations and they sum up to a 3-D image of \numprint{32} channels. We only took the center slice from each channel, as there is no variation for objects in the vertical direction. Thus, the input data of each object is a 32-channel 2-D image. The output data is a 1-channel binary image indicating object occupancy in the horizontal plane. The simulation for \numprint{15750} training objects took \numprint{20} days on a Tesla P100 GPU.

An example of our input-output pairs is shown in \cref{fig: input-output pair}. On the left-hand side are \numprint{32} image channels representing simulations of loudness fields for \numprint{4} frequency ranges and \numprint{8} source directions. The black region at the center of an image channel denotes region $\mathcal{R}$, which contains an object. Pixels for $\mathcal{S}$ are associated with numerical values in decibels. On the right-hand side is an binary image representing object occupancy in $\mathcal{R}$. Black pixels of value \numprint{0} denote air occupancy and white pixels of value \numprint{1} denote object occupancy.

\subsection{Resolution degradation and channel reduction}
Acquisition of input images in \cref{fig: input-output pair} is  expensive, considering the spatial density of probes and the numbers of source locations and energy bands. Thus, we aim to study the robustness of our approach to the degradation of image resolution and the reduction of image channels. Our frequency ranges are divided into a low group and a high group, each of \numprint{2} octave bands. Using either one of them or both leads to \numprint{3} situations. Further, using half of the source locations or using all of them leads to \numprint{2} situations. Finally, we define \numprint{4} spatial sampling factors: \numprint{1}, \numprint{2}, \numprint{4} and \numprint{8}, which correspond to situations of no sampling, sampling every \numprint{1} in \numprint{2} pixels, every \numprint{1} in \numprint{4} pixels and every \numprint{1} in \numprint{8} pixels in \numprint{2} dimensions. All the above situations lead to \numprint{24} combinations of image channels and resolutions. An example of the dataset using all \numprint{8} source locations, low octave bands, and a spatial sampling factor of 2 is shown in \cref{fig: decimated input-output pair}. 

\begin{table}[t!]
\caption{IMED \cite{wang2005euclidean} between predictions and ground truth, normalized between \numprint{0} and \numprint{1}. SSF is short for spatial sampling factor.}
\centering
\noindent
\begin{tabular}{N N N N N N N }\toprule
\multicolumn{1}{N }{\textbf{}} & \multicolumn{3}{c }{\textbf{4 Sources}} & \multicolumn{3}{c }{\textbf{8 Sources}} \\  

\cmidrule(lr){2-4}
\cmidrule(ll){5-7}
\multicolumn{1}{N }{\textbf{}} & \multicolumn{3}{c }{\textbf{Frequency Band}} & \multicolumn{3}{c }{\textbf{Frequency Band}} \\ 
\multicolumn{1}{N }{\textbf{SSF}} & \textbf{Low} & \textbf{High} & \textbf{Full} & \textbf{Low} & \textbf{High} & \textbf{Full} \\ 
\cmidrule(lr){1-1}
\cmidrule(lr){2-4}
\cmidrule(ll){5-7}
\multicolumn{1}{ c }{8} & 0.1453 & 0.1819 & 0.1844 & 0.1420 & 0.1993 & 0.1680  \\ 
\multicolumn{1}{ c }{4} & 0.1491 & 0.2136 & 0.1753 & 0.1524 & 0.1881 & 0.1610  \\ 
\multicolumn{1}{ c }{2} & 0.1603 & 0.2036 & 0.1677 & 0.1499 & 0.1880 & 0.1732  \\ 
\multicolumn{1}{ c }{1} & 0.1497 & 0.2004 & 0.1600 & 0.1480 & 0.1893 & 0.1679  \\ 
\bottomrule
\end{tabular}
\label{tab: distance}
\end{table}

\begin{figure}[t!]
\centering
\includegraphics[width=0.45\textwidth]{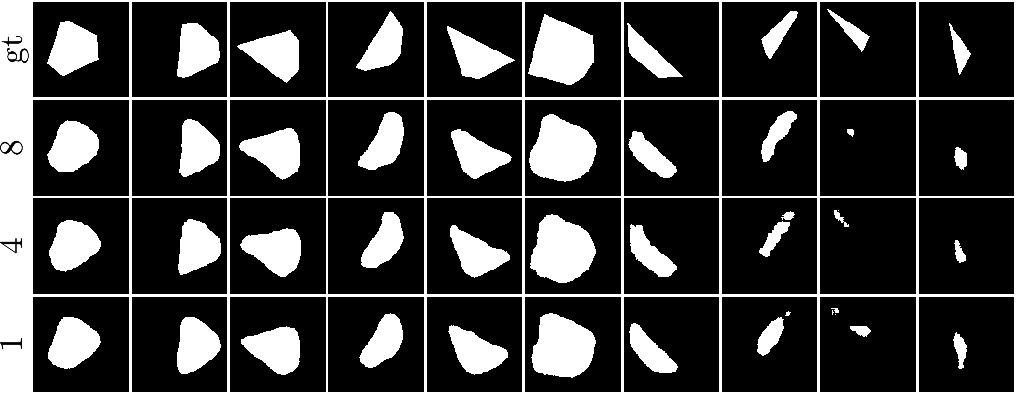}
\caption{Comparison of predictions using different spatial subsapmling factors. Notation ``gt'' is short for ``ground truth''. Predictions using spatial sapmling factors \numprint{8}, \numprint{4}, \numprint{1} are shown in separate rows. Source numbers are uniformly \numprint{8} and only low octave bands are used.}
\label{fig: samp comparison}
\vspace*{\floatsep}
\includegraphics[width=0.45\textwidth]{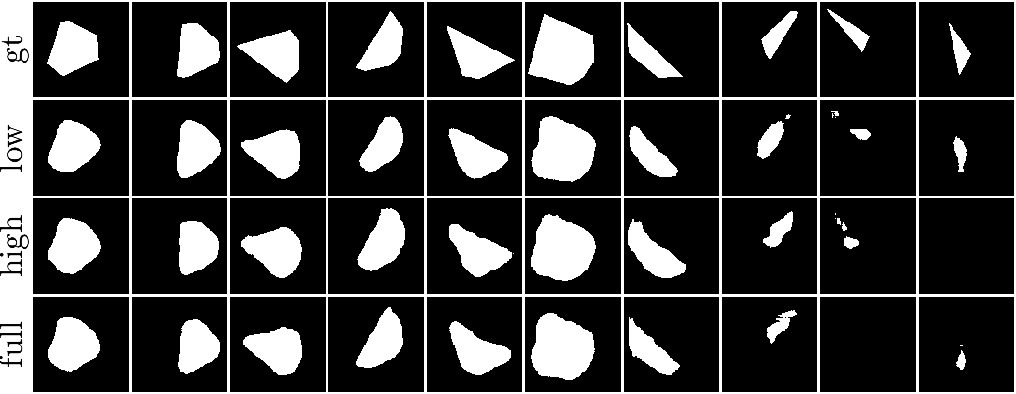}
\caption{Comparison of predictions using different combinations of octave bands. Predictions using lower octave bands, higher octave bands and both are illustrated in different rows. Numbers of sources are uniformly \numprint{8} and spatial sampling factors are uniformly \numprint{1}.}
\label{fig: freq comparison}
\vspace*{\floatsep}
\includegraphics[width=0.45\textwidth]{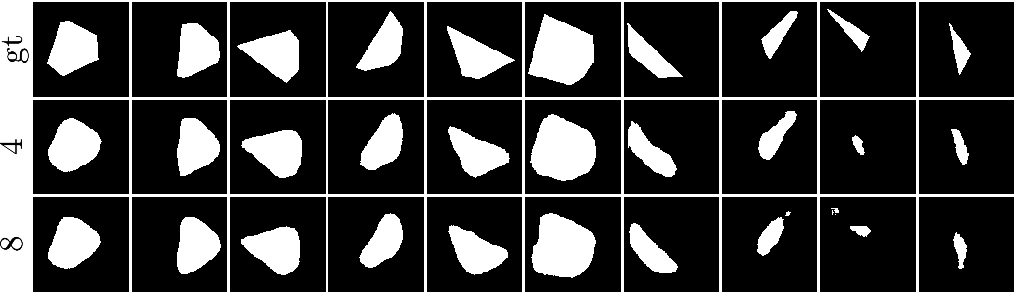}
\caption{Comparison of predictions using different numbers of sources. Predictions using \numprint{4} and \numprint{8} source channels are illustrated in different rows. Spatial sampling factors are uniformly \numprint{1} and only low octave bands are used.}
\label{fig: source comparison}
\end{figure}

\subsection{Convolutional neural network}
We used the full resolution residual network (FRRN) \cite{Pohlen2017Frrn} since it has been shown to successfully learn the mapping from geometry to loudness\cite{Fan2019Fast}, adopting the implementation from Shah \cite{mshahsemseg}. We kept the architecture of the FRRN unchanged from Pohlen's work \cite{Pohlen2017Frrn} and modified the numbers of input and output channels to our needs. Cross-entropy was used as the loss metric. Targets were selected as the binary images of size \numprint{100}$\times$\numprint{100} denoting the inaccessible region. We also adopted mini-batches of size 2, selected stochastic gradient descent as our optimizer and set the learning rate, momentum, and weight decay as $1\mathrm{e}{-5}$, $0.99$ and $1.0\mathrm{e}{-5}$ respectively. Combinations of spatial sampling factors, octave bands, and source locations were differentiated in the data loader. We trained \numprint{24} networks corresponding to the \numprint{24} datasets on a GPU cluster. Each network was trained for at least \numprint{4} epochs. Training took a total of \numprint{4} days.

\subsection{Generalization test}
Our test data was processed in the same manner as the training data. As a result, we obtained \numprint{24} test datasets for the \numprint{24} CNN models. Test input images were fed into their corresponding models and each model output \numprint{700} predictions for geometry of test objects. Examples of the geometric predictions are shown in \cref{fig: samp comparison}, \cref{fig: freq comparison} and \cref{fig: source comparison}.

For the comparison in \cref{fig: samp comparison}, choices of octave bands and number of sources are kept the same and the only variable is the spatial sampling factor. In general, severe degradation is not observed when a coarser spatial resolution is used, as shown by a comparison between row \numprint{2}, \numprint{3} and \numprint{4}. A direct comparison between the predictions and the ground truth reveals that sharp corners in ground-truth objects are rounded in predictions. This discrepancy is induced by the smoothing effect of CNNs. It should also be noted that the predictions of smaller objects are not as accurate as larger ones, as shown by a comparison between columns \numprint{8}, \numprint{9}, and \numprint{10} and all previous columns. Specifically, only small portions of the triangles in columns \numprint{9} and \numprint{10} are predicted and the predicted object boundaries diverge from the ground truth. Similar tendencies can also be found in \cref{fig: freq comparison} and \cref{fig: source comparison}, where the variables for comparison are respectively frequency ranges and source distributions.

The IMage Euclidean Distance (IMED) \cite{wang2005euclidean} is used to quantify prediction accuracy of the \numprint{24} models. Surprisingly, models using low octave bands are superior to models using high octave bands or both, as shown in \cref{tab: distance}. Diffraction is more prominent at lower frequencies, and acoustic waves tend to ``wrap'' around objects as diffraction takes place. Thus, one potential explanation for the superiority of using low octave bands is, the ``wrapping'' effect of diffraction helps probe a whole surface, while only partial local features are revealed by refection and occlusion, which are more prominent at higher frequencies. Further, using coarser images or fewer source locations does not significantly degrade prediction accuracy. This finding is consistent with direct observation from \cref{fig: samp comparison} and \cref{fig: source comparison}. The only model showing decreased accuracy with spatial sampling is the one trained with \numprint{4} source locations and \numprint{2} frequency bands. Thus, our CNN models are robust to reduction of both source channels and spatial resolution.

%% file: text/Conclusion.tex
\section{conclusion}
In this study, inference of object geometry is formulated as image segmentation. Training and test data are generated from a fast acoustic solver. CNN models are shown to predict object geometry with high accuracy. Further, prediction robustness to reduction of acoustic image channels and of spatial resolution is validated. Specifically, models trained using low-frequency images are superior to models trained using high-frequency images or both, indicating the significance of diffraction in the inference.

There is still much to improve in our work. Only one object is considered in our inference and our objects are convex horizontally and do not vary vertically. These simplifications restrict the generalizability of our work. In the future, we aim to infer the geometry of multiple prisms, lifting the restriction on convexity. Further, no background noise is assumed. In a future work, reflections from the ground should be included in datasets. Finally, our models are trained and tested using only simulations, and it would be interesting to examine the robustness of our approach to real measurements.

\clearpage

%% file: main.bbl
\begin{thebibliography}{10}
\providecommand{\url}[1]{#1}
\def\UrlFont{\rmfamily}
\providecommand{\newblock}{\relax}
\providecommand{\bibinfo}[2]{#2}
\providecommand\BIBentrySTDinterwordspacing{\spaceskip=0pt\relax}
\providecommand\BIBentryALTinterwordstretchfactor{4}
\providecommand\BIBentryALTinterwordspacing{\spaceskip=\fontdimen2\font plus
\BIBentryALTinterwordstretchfactor\fontdimen3\font minus
  \fontdimen4\font\relax}
\providecommand\BIBforeignlanguage[2]{{%
\expandafter\ifx\csname l@#1\endcsname\relax
\typeout{** WARNING: IEEEtran.bst: No hyphenation pattern has been}%
\typeout{** loaded for the language `#1'. Using the pattern for}%
\typeout{** the default language instead.}%
\else
\language=\csname l@#1\endcsname
\fi
#2}}

\bibitem{Antonacci2010Reconstruction}
F.~{Antonacci}, A.~{Sarti}, and S.~{Tubaro}, ``Geometric reconstruction of the
  environment from its response to multiple acoustic emissions,'' in \emph{2010
  IEEE International Conference on Acoustics, Speech and Signal Processing},
  March 2010, pp. 2822--2825.

\bibitem{Ba2010L1}
D.~{Ba}, F.~{Ribeiro}, C.~{Zhang}, and D.~{Florêncio}, ``L1 regularized room
  modeling with compact microphone arrays,'' in \emph{2010 IEEE International
  Conference on Acoustics, Speech and Signal Processing}, March 2010, pp.
  157--160.

\bibitem{Baba2016Multiple}
Y.~E. {Baba}, A.~{Walther}, and E.~A.~P. {Habets}, ``Reflector localization
  based on multiple reflection points,'' in \emph{2016 24th European Signal
  Processing Conference (EUSIPCO)}, Aug 2016, pp. 1458--1462.

\bibitem{Canclini2011Doa2Loc}
A.~{Canclini}, P.~{Annibale}, F.~{Antonacci}, A.~{Sarti}, R.~{Rabenstein}, and
  S.~{Tubaro}, ``From direction of arrival estimates to localization of planar
  reflectors in a two dimensional geometry,'' in \emph{2011 IEEE International
  Conference on Acoustics, Speech and Signal Processing (ICASSP)}, May 2011,
  pp. 2620--2623.

\bibitem{Canclini2011Exact}
A.~{Canclini}, F.~{Antonacci}, M.~R.~P. {Thomas}, J.~{Filos}, A.~{Sarti}, P.~A.
  {Naylor}, and S.~{Tubaro}, ``Exact localization of acoustic reflectors from
  quadratic constraints,'' in \emph{2011 IEEE Workshop on Applications of
  Signal Processing to Audio and Acoustics (WASPAA)}, Oct 2011, pp. 17--20.

\bibitem{Dokmanic2013NAS}
\BIBentryALTinterwordspacing
I.~Dokmani{\'c}, R.~Parhizkar, A.~Walther, Y.~M. Lu, and M.~Vetterli,
  ``Acoustic echoes reveal room shape,'' \emph{Proceedings of the National
  Academy of Sciences}, vol. 110, no.~30, pp. 12\,186--12\,191, 2013.
\BIBentrySTDinterwordspacing

\bibitem{Dokmanic2011RoomShape2D}
I.~{Dokmanić}, Y.~M. {Lu}, and M.~{Vetterli}, ``Can one hear the shape of a
  room: The 2-d polygonal case,'' in \emph{2011 IEEE International Conference
  on Acoustics, Speech and Signal Processing (ICASSP)}, May 2011, pp. 321--324.

\bibitem{Filos2012Linear}
J.~{Filos}, A.~{Canclini}, F.~{Antonacci}, A.~{Sarti}, and P.~A. {Naylor},
  ``Localization of planar acoustic reflectors from the combination of linear
  estimates,'' in \emph{2012 Proceedings of the 20th European Signal Processing
  Conference (EUSIPCO)}, Aug 2012, pp. 1019--1023.

\bibitem{Filos2010TwoStep}
J.~Filos, E.~A. Habets, and P.~A. Naylor, ``A two-step approach to blindly
  infer room geometries,'' in \emph{Proc. Int. Workshop Acoust. Echo Noise
  Control (IWAENC)}.\hskip 1em plus 0.5em minus 0.4em\relax Citeseer, 2010.

\bibitem{Michael2019Spherical}
\BIBentryALTinterwordspacing
M.~Lovedee-Turner and D.~Murphy, ``Three-dimensional reflector localisation and
  room geometry estimation using a spherical microphone array,'' \emph{The
  Journal of the Acoustical Society of America}, vol. 146, no.~5, pp.
  3339--3352, 2019.
\BIBentrySTDinterwordspacing

\bibitem{Moore2013SingleChannel}
A.~H. {Moore}, M.~{Brookes}, and P.~A. {Naylor}, ``Room geometry estimation
  from a single channel acoustic impulse response,'' in \emph{21st European
  Signal Processing Conference (EUSIPCO 2013)}, Sep. 2013, pp. 1--5.

\bibitem{Naseri2017Cooperative}
H.~{Naseri} and V.~{Koivunen}, ``Cooperative simultaneous localization and
  mapping by exploiting multipath propagation,'' \emph{IEEE Transactions on
  Signal Processing}, vol.~65, no.~1, pp. 200--211, Jan 2017.

\bibitem{Nastasia2011Localization}
E.~{Nastasia}, F.~{Antonacci}, A.~{Sarti}, and S.~{Tubaro}, ``Localization of
  planar acoustic reflectors through emission of controlled stimuli,'' in
  \emph{2011 19th European Signal Processing Conference}, Aug 2011, pp.
  156--160.

\bibitem{Remaggi2014Room}
L.~{Remaggi}, P.~J.~B. {Jackson}, P.~{Coleman}, and W.~{Wang}, ``Room boundary
  estimation from acoustic room impulse responses,'' in \emph{2014 Sensor
  Signal Processing for Defence (SSPD)}, Sep. 2014, pp. 1--5.

\bibitem{Remaggi2017Novel}
L.~{Remaggi}, P.~J.~B. {Jackson}, P.~{Coleman}, and W.~{Wang}, ``Acoustic
  reflector localization: Novel image source reversion and direct localization
  methods,'' \emph{IEEE/ACM Transactions on Audio, Speech, and Language
  Processing}, vol.~25, no.~2, pp. 296--309, Feb 2017.

\bibitem{Remaggi20153D}
L.~{Remaggi}, P.~J.~B. {Jackson}, W.~{Wang}, and J.~A. {Chambers}, ``A 3d model
  for room boundary estimation,'' in \emph{2015 IEEE International Conference
  on Acoustics, Speech and Signal Processing (ICASSP)}, April 2015, pp.
  514--518.

\bibitem{Remaggi2015Source}
L.~Remaggi, P.~J. Jackson, and P.~Coleman, ``Source, sensor and reflector
  position estimation from acoustical room impulse responses,'' \emph{Proc.
  Int. Congr. Sound Vibration}, pp. 1--8, 2015.

\bibitem{Tervo20123D}
S.~{Tervo} and T.~{Tossavainen}, ``3{D} room geometry estimation from measured
  impulse responses,'' in \emph{2012 IEEE International Conference on
  Acoustics, Speech and Signal Processing (ICASSP)}, March 2012, pp. 513--516.

\bibitem{Antonacci2012InferenceRoom}
F.~{Antonacci}, J.~{Filos}, M.~R.~P. {Thomas}, E.~A.~P. {Habets}, A.~{Sarti},
  P.~A. {Naylor}, and S.~{Tubaro}, ``Inference of room geometry from acoustic
  impulse responses,'' \emph{IEEE Transactions on Audio, Speech, and Language
  Processing}, vol.~20, no.~10, pp. 2683--2695, Dec 2012.

\bibitem{Dokmanic2016RoomSlam}
I.~{Dokmani\'{c}}, L.~{Daudet}, and M.~{Vetterli}, ``From acoustic room
  reconstruction to slam,'' in \emph{2016 IEEE International Conference on
  Acoustics, Speech and Signal Processing (ICASSP)}, March 2016, pp.
  6345--6349.

\bibitem{Lindell2019Nlos}
D.~B. Lindell, G.~Wetzstein, and V.~Koltun, ``Acoustic non-line-of-sight
  imaging,'' in \emph{The IEEE Conference on Computer Vision and Pattern
  Recognition (CVPR)}, June 2019.

\bibitem{Fan2019Fast}
Z.~{Fan}, V.~{Vineet}, H.~{Gamper}, and N.~{Raghuvanshi}, ``Fast acoustic
  scattering using convolutional neural networks,'' in \emph{2020 IEEE
  International Conference on Acoustics, Speech and Signal Processing
  (ICASSP)}, 2020, pp. 171--175.

\bibitem{williams1999fourier}
E.~G. Williams, \emph{Fourier acoustics: sound radiation and nearfield
  acoustical holography}.\hskip 1em plus 0.5em minus 0.4em\relax Elsevier,
  1999.

\bibitem{fan2019computation}
\BIBentryALTinterwordspacing
Z.~Fan, T.~Arce, C.~Lu, K.~Zhang, T.~W. Wu, and K.~McMullen, ``Computation of
  head-related transfer functions using graphics processing units and a
  pereptual validation of the computed hrtfs against measured {HRTF}s,'' in
  \emph{Audio Engineering Society Conference: 2019 AES International Conference
  on Headphone Technology}, Aug 2019.
\BIBentrySTDinterwordspacing

\bibitem{wu2002boundary}
T.~Wu, ``Boundary element acoustics fundamentals and computer codes,'' 2002.

\bibitem{raghuvanshi2018triton}
\BIBentryALTinterwordspacing
N.~Raghuvanshi and J.~Snyder, ``Parametric directional coding for precomputed
  sound propagation,'' \emph{ACM Trans. Graph.}, vol.~37, no.~4, pp.
  108:1--108:14, July 2018.
\BIBentrySTDinterwordspacing

\bibitem{wang2005euclidean}
L.~Wang, Y.~Zhang, and J.~Feng, ``On the euclidean distance of images,''
  \emph{IEEE transactions on pattern analysis and machine intelligence},
  vol.~27, no.~8, pp. 1334--1339, 2005.

\bibitem{Pohlen2017Frrn}
T.~Pohlen, A.~Hermans, M.~Mathias, and B.~Leibe, ``Full-resolution residual
  networks for semantic segmentation in street scenes,'' in \emph{The IEEE
  Conference on Computer Vision and Pattern Recognition (CVPR)}, July 2017.

\bibitem{mshahsemseg}
M.~P. Shah, ``Semantic segmentation architectures implemented in pytorch,''
  \url{https://github.com/meetshah1995/pytorch-semseg/}, 2017, accessed: June,
  2019.

\end{thebibliography}
